\documentclass[oldversion]{aa} 
\usepackage{lscape}
\usepackage{graphicx}
\usepackage{txfonts}

\newcommand{\s}{SnIa }
\newcommand{\be}{\begin{equation}}
\newcommand{\ee}{\end{equation}}

\begin{document}
\title{Redshift--distance relations from type Ia supernova observations}
\subtitle{New constraints on grey dust models}
\author{Aday R. Robaina\inst{1,2}\fnmsep\thanks{Currently
at the Max-Planck-Institut f\"ur Astronomie, Heidelberg, Germany. arobaina@mpia-hd.mpg.de}
\and
 Jordi Cepa\inst{2,3}
} 

\institute{Facultad de F\'\i sica, Universidad de La Laguna E-38071, La Laguna, Tenerife, Spain \\
\email{arobaina@iac.es}
\and
Instituto de Astrof\'\i sica de Canarias E-38200 La Laguna, Tenerife, Spain
\and
Departamento de Astrof\'\i sica, Facultad de F\'\i sica, Universidad de La Laguna. 
E-38071 La Laguna, Tenerife, Spain\\
 \email{jcn@iac.es}
}
\date{Received 28 March 2006; accepted 03 August 2006}

\abstract
{Extinction due to intergalactic grey dust has been proposed as an alternative to 
 accelerated expansion to account for the dimming of \s fluxes beyond $z \simeq 0.5$. 
The ``replenishing'' grey dust model, although fitting the observational data, does not seem to be 
based on physical assumptions. For this reason, in this paper we propose a new grey dust model whose 
dust distribution follows the comoving SFR density evolution, a reasonably
 established phenomenon. 
This new model is compared with the updated photometric data sample from the High 
Z Supernova Search Team (HZT). Also, using pairs of supernovae at different redshifts, 
the possibility of any ``patchy'' distribution proposed to explain the discrepancies between the 
``high $z$'' dust model and the \s observations at very high redshift ($z \ge 0.9$) is ruled out. 
Finally, the data are compared with different models with and without dark energy 
and a best fit to a universe with adimensional parameters  $\Omega_{\mathrm{m0}} = 0.31$ and 
$\Omega_{\mathrm{\Lambda 0}} = 0.69$ is obtained, from which an age of the universe of
$t_{0} \simeq 14.6 \times 10^{9}$ years is derived. This age is compatible with the age 
of the globular clusters using an equation of state $\omega = -1$ and it obviates the need to
resort to any kind of phantom energy.} 

\maketitle
%

\section{Introduction}
After the discovery by Riess et al.(1998) and Perlmutter et al.(1999) of the excessive weakness of type 
Ia supernovae (\s) at high redshift, cosmology has undergone a revolution similar to the cosmic microwave background (CMB) discovery. The cosmological interpretation of these observations 
led to the postulation of the accelerated expansion of the universe as a consequence of a ``dark energy'' of 
negative pressure and unknown origin. Given the difficult physical interpretation of this dark energy,
either in terms of a cosmological constant,  quintessence, or even some form of phantom energy, several mechanisms 
have been proposed to explain \s observations without resorting to an accelerated expansion. Among 
them are models of intergalactic grey dust, which try to explain the dimming of the observed 
brightness at $z \simeq 0.5$ by assuming that an absorbent medium exists in
intergalactic space 
whose extinction does not depend on wavelength. These dust models have been driven to evolve at the same 
rate as the data supplied by  \s teams. 

The first models, proposed by Aguirre (1999a, 1999b), were based on both a model of dust formed by carbon 
needles and a Draine \& Lee dust model (Draine \& Lee 1984), obtaining a density parameter of 
$\Omega_{\mathrm{Dust}}\sim 10^{-5}$. Also, Goobar, Bergstrom \& Mortsell (2002) have developed a 
model of intergalactic dust whose effects on the distance modulus vs.\ $z$ are indistinguishable 
from those caused by a $\Lambda$CDM universe with $\Omega_{\mathrm{m0}}\simeq0.3, \Omega_{\mathrm{\Lambda
0}}\simeq0.7$ (more information on the intergalactic medium amount and its effects can be found in, for
example,
Bianchi \& Ferrara 2005; Inoue \& Kamaya 2003; Nath, Sethi \& Shchekinov 1999).

In any case, grey dust models imply universes with $\Omega_{\mathrm{m0}}=1$ or non-Euclidean geometries. 
The first possibility can be discarded by the 2dF experiment (Peacock et al.\ 2001), while the second can be 
rejected by analysis of the anisotropies observed in the CMB (Spergel et al.\ 2003), from which a total parameter $\Omega=1$ is derived.

Apart from grey dust models, another effect that could hamper the need for an accelerated expansion universe 
to explain \s data is their luminosity evolution. Observational evidence shows \s luminosity variations 
with galactic morphology (Hamuy et al.\ 1996, 2000; Branch et al.\ 1996; Reindl et al.\ 2005), while 
some theoretical studies (Hoflich et al.\ 1998, 2000) suggest that \s luminosity could have some 
dependence on redshift. Were this true, this would question the status of standard candle 
for this kind of objects. In any case, these studies are beyond the scope of this article and are merely
mentioned  as further possibilities in addition to grey dust models in order to avoid accelerated
expansion.

In this paper we present observational evidence to refute the high-$z$ grey dust model even in the
case of patchy distribution. We also propose a new grey dust model with a density distribution 
following that of the star formation rate (SFR) density. This SFR-based distribution is based on 
observational data, thus making the grey dust model based on physical assumptions that best fits 
the supernova data. In Sect. 2 we give an overview of  existing theoretical grey dust models, 
including  high-$z$ and the replenishing grey dust. In this section the new SFR-based 
distribution model is also developed . In Sect. 3 we derive the dust-free
model by best-fitting the \s data, 
obtaining its density parameters and the age of the universe, and compare the data with  dust models, 
including the one developed here. We also explore the possibility of reconciling the observational data 
with the high-$z$ dust model by means of a patchy dust distribution. Finally, we draw our conclusions 
 in Sect. 4.

\section{Theoretical models}

The existence of grey dust would have an enormous impact on the analysis of
the High Z Supernova Search Team 
(HZT) and the Supernova Cosmology Project (SCP) observations, returning the standard model of the universe 
to the point where it was before 1998 with their contradictions of the CMB anisotropy experiments, which 
point to a flat universe ($\Omega=1$). This is the reason a detailed study of the different possibilities
is required.

Any intergalactic dust model producing  grey extinction relevant to the study of \s observations must
have the following characteristics:

\begin{enumerate}
\item Large grain size, to account for the achromaticy of the absortion and to comply with the 
mechanism of preferential destruction of smaller size grains (Aguirre 1999b).
\item An efficient expulsion mechanism in the galaxy where the grains are formed. It must explain  
 why only this type of dust is found in the IGM.
\item Coherence with the expected metallicity of the intergalactic medium.
\item The dust distribution must be based on a hypothesis in turn based on reasonably established physical observables.
\end{enumerate}

\subsection{High-z grey dust}

Even though Aguirre's (1999a, 1999b) high-$z$ dust model implies a uniform dust distribution 
similar to that of non-relativistic matter,

\be
\rho_{\rm dust}=\rho_{\rm dust}^{0}(1+z)^{3},
\ee
which fulfils the first three   of the four requirements mentioned above, the latest results published by the HZT (Riess et al.\ 2004), 
with \s data beyond $z=1.5$,  almost certainly (at the $4 \sigma$ level) reject the possibility of this 
high-$z$ dust model. Figure 1 shows the large deviation of the Aguirre (1999a, 1999b) model in the 
high-$z$ region compared to the experimental data and  the best fit.

However a patchy dust distribution could render this conclusion unsound. If the intergalactic dust 
distribution were not homogeneous, it might be possible that the light coming from an \s at high redshift 
($z \ge 0.9$) propagates through more or less absorbent medium, depending on the source position 
in the sky. The deviation of the observed data from the high-$z$ dust model would then be due to a bias 
produced by the grey dust, leaving the magnitude of the absorbed supernovae below the detection limit 
of  current telescopes. In this case, the sample observed at very high redshift would belong to the 
group of \s whose optical path has not been very affected by the aforementioned absortion, and for this reason 
its apparent brightness would be higher than that inferred by supposing a completely homogeneous dust 
distribution.

The inhomogeneity in the grey dust distribution is a solid enough possibility to take it into account 
in spite of Aguirre's (1999a, 1999b) arguments. In his theoretical study, possible mechanisms for preferential 
ejection of grey dust from the galaxies and the temporal scale on which it could have been homogeneously 
distributed, are explained. They are partly corroborated by the work of Renzini (1997) on the metallicity 
of the intracumular gas. Nevertheless, it is difficult to explain how a dust distribution formed by 
baryonic matter can escape from the baryonic large-scale structures observed in the universe. Aguirre 
himself (1999a) suggests that the study of systematic differences between the flux received from \s hosted
by field and cluster galaxies could distinguish between a universe in accelerated expansion and an effect 
due to grey dust, respectively, by restricting implicitly the homogenity to the intracumular space.

\subsection{Replenishing grey dust}

As different surveys  reached more and more distant redshifts, it  became evident that the high-$z$ 
dust model was unable to explain the data. However, the reasonable values obtained by Aguirre (1999a, 1999b) 
for the time scale of the expulsion mechanisms of his grain models have promoted the birth of new models 
that fit the data much better. A good example is the replenishing grey dust model, proposed by 
Goobar, Bergstrom \& Mortsell (2002), whose match to the data is at least as good as the 
$\Omega_{\Lambda} \ne 0$ models.

This replenishing grey dust model is based on a grey dust extinction that follows a density distribution
of the form:

\be
\rho_{\rm dust}=\rho_{\rm dust}^{o}(1+z)^{a},
\ee
where 

\begin{displaymath}
a = \left\{ \begin{array}{ll}

 3 & \textrm{if $z < 0.5$}\\
 0 & \textrm{if $z > 0.5$.}
  \end{array} \right.
\end{displaymath}
This model requires  extremely efficient formation and expulsion  of the dust in the range $z > 0.5$
so that the dilution due to the expansion of the universe is exactly compensated by the metal formation 
in galaxies and its further removal. However, this behaviour is not consistent with the current knowledge of
the overall star formation episodes that took place in the universe. As a consequence, a physical 
explanation of this replenishing mechanism is still lacking. Moreover, the hitherto unjustified  sharp transition 
from $a=0$ to $a=3$ represents a dramatic end of metal formation and expulsion from the galaxy without a
transition era.

\subsection{SFR-based model}

To avoid these features of the replenishing dust model, a grey dust model based on the 
star formation rate (SFR) density evolution is proposed here. It will be shown that the
proposed model fits  the existing \s data equally well, while at the same time providing a dust production based on the
current knowledge of  star formation evolution in the universe.

\begin{figure}[h!]
\begin{center}
\includegraphics[width=8cm,height=6.5cm]{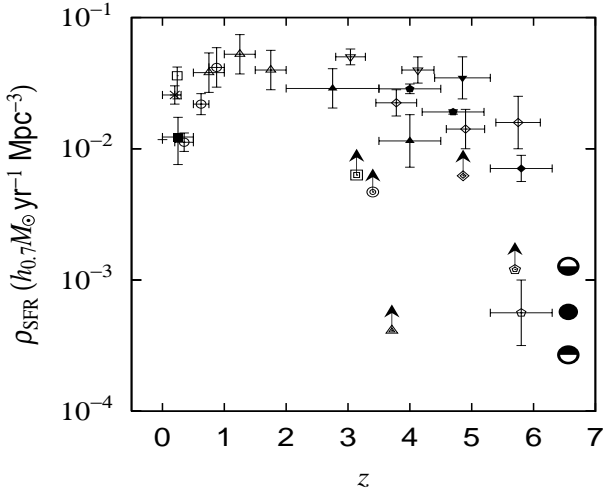}
\label{Figure 1}
\caption{\footnotesize{Comoving $\rho_{\mathrm{SFR}}$ as function of  redshift. Figure taken from 
Taniguchi et al.\ (2005)}}
\end{center}
\end{figure}

Following the results compiled by Taniguchi et al.\ (2005) on the comoving SFR density evolution 
(Figure 2), it can be observed that the comoving SFR density ($\rho^{\mathrm{c}}_{\mathrm{SFR}}$) is 
approximately constant in the range $1 \le z \le 2$. In the region $0 < z < 1$, the SFR density 
may be reasonably represented by a simple relation with redshift, with lower SFR at recent times. 
For simpler description, the SFR density curve is then represented by these two regimes. For our
 purposes this simple parameterization is adequate, but see Oda \& Totani 2005 for more complex parameterization.

  In this context, the comoving SFR density is represented by

\begin{displaymath}
\rho^{\mathrm{c}}_{\mathrm{SFR}}\varpropto \left\{ \begin{array}{ll}

 10^{1+z} & \textrm{if \hspace{0.7cm}$0 < z < 1$}\\
 \textrm {constant}  & \textrm{if \hspace{0.7cm}$1 < z < 2$.}\\
  \end{array} \right.
\end{displaymath}
The evolution of the SFR proper density ($\rho^{\mathrm{p}}_{\mathrm{SFR}}$) is

\be
\rho_{\mathrm{SFR}}^{\mathrm{p}}(z) \varpropto \int_{z_{i}}^{z} \frac{\rho^{\mathrm{c}}_{\mathrm{SFR}} 
\cdot (1+z)^{1/2}}{a_{0}^{3} \cdot H_{0}} dz,
\ee
where $a_{0}$ is the scale factor at present and the value taken for $H_{0}$ is 65 km s$^{-1}$ Mpc$^{-1}$.

Nevertheless, although the dust density derives from the SFR evolution, not all the dust escapes from
the galaxy to the IGM. At least a fraction of the metals produced will be retained in low mass stars 
that never reach the core collapse supernova phase and do not present strong stellar winds. An estimate 
of the retained metal fraction can be obtained from the Salpeter initial mass function (Salpeter 1955) 
from $0.1\ M_{\odot}$ to $8\ M_{\odot}$. This SFR-based distribution will then include, as explained above,
a transition at $z \simeq 1$ between the two regimes of maximum and decreasing SFR:

\begin{displaymath}
\rho_{\mathrm{dust}} \varpropto \left \{ \begin{array}{ll}

 \rho_{\mathrm{dust1}}^{0} \Big( - \frac{\sqrt{\pi}}{2\ln{10}^{3/2}} Erfi \big(\sqrt{1+z} \sqrt{\ln{10}}\big)+\frac{10^{1+z} \sqrt{1+z}}{\ln{10}} \Big)  
       & \textrm{if $z < 1$}\\
 \rho_{\mathrm{dust2}}^{0}  \big( \frac{2}{3} (1+z)^{3/2}\big) & \textrm{if $z > 1$.}\\
  \end{array} \right.
\end{displaymath}
The constants $\rho_{\mathrm{dust1}}^{0}$, and $\rho_{\mathrm{dust2}}^{0}$, which include the estimation 
of metals retained in low mass stars, must be fitted in the region ($0<z<1$) to match the expected 
deviation at $z=0.5$, while they are chosen to guarantee the continuity of the
curve in the region $z>1$.

Although the dust is produced by the star formation, the relevant physical magnitude to evaluate 
is the opacity, not the SFR. The resulting extinction is then delayed with respect to dust production 
owing to the time interval elapsed from production to insertion into the IGM. This delay can be estimated 
from assumptions for the velocity of the ejecta.

In order to fulfil the above-mentioned requirements, this grey dust model must 
present grain characteristics similar to those proposed in previous models, reproducing the expected 
values for $R_{v}$ and absorptions (Goobar, Bergstrom \& Mortsell 2002). Finally, the injection 
mechanism and the preferential destruction of small size grains will be assumed to be those 
discussed by Aguirre (1999b, 2001).

\section{Results}

\subsection{Cosmological constraints}

In this work we have used the latest published HZT data  (Riess et al.\ 2004; Strolger et al.\ 2004), since 
it represents the most significant sample with photometric values to date. This sample has greatly benefited 
from the results of the {Great Observatories Origins Deep Survey} (GOODS) project, carried out 
using the {Advanced Camera for Surveys} (ACS) on board the \emph{Hubble Space
Telescope} (\emph{HST}). 
The \s classification in the groups Gold, Silver, and Bronze used is explained in full detail by 
Strolger et al.\ (2004). The most certain classification is the Gold class, characterised by a clear 
spectroscopic confirmation of the supernova type. 

In this section, the observational data are fitted by a model composed of 
non-relativistic matter and dark energy, neglecting the contributions of radiation and relativistic 
matter, which do not play a significant role in the universe in the redshift range considered. 
The magnitude used for the fit is based on the distance modulus,

\be
m-M=5 \log D_{\mathrm{L}} +25,
\ee
where $m$ and $M$ are the apparent and absolute magnitudes, respectively, and $D_\mathrm{L}$ is the
luminosity distance in Mpc. Assuming a flat geometry, as derived from the spectra of the CMB
anisotropies, we have obtained, via an $\chi^{2}$ criterion, the best fit with the parameters 
$\Omega_{\mathrm{m0}}=0.31$ and $\Omega_{\mathrm{\Lambda 0}}=0.69 $, values similar to those found in 
previous studies (Riess et al.\ 1998, 2004; Perlmutter et al.\ 1999), and consistent with those derived
from 2dF results (Peacock et al.\ 2001).   

These values are only indicative, given the fit dependence with the rejecting 
criterion. In Figure 2 we show the Hubble diagram of the entire sample, together with the best fit found. 

\begin{figure}[h!]
\begin{center}
\includegraphics[width=9cm,height=7.5cm]{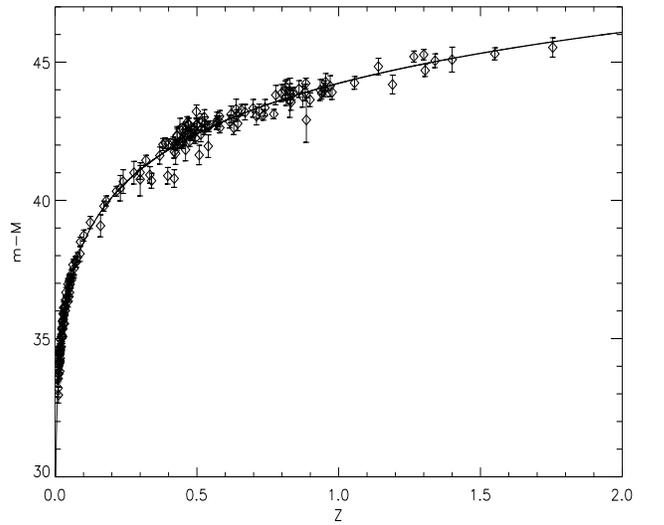}
\caption{\footnotesize{ Hubble diagram. Distance moduli of observational \s data $vs$ $z$. Overplotted  
 are the  $\Omega_{\mathrm{m0}}= 0.31$, $\Omega_{\mathrm{\Lambda 0}}= 0.69$ universe, which is the 
best fit for a flat cosmology obtained from $\chi^{2}$ statistics.}}
\end{center}
\end{figure}

It is worth noting that these values of mass and dark-energy density parameters, assuming a 
Hubble constant of 65 km s$^{-1}$ Mpc$^{-1}$, yield an age of the universe around $t_0\simeq 14.6\times
10^{9}$ years as derived using Cepa (2004), with the constraint $\omega = -1$ for the equation of 
state of dark energy (i.e.\ a cosmological constant). This result is very nearly compatible with the 
globular cluster age of $12.9 \pm 2.9$ Gy found by Carretta et al.\ (2000) and its time of formation, 
without resorting to a phantom energy (Cepa 2004), although the error margin is still too large to 
determine the equation of state of the dark energy.

\subsection{The new dust model}

The \s weakness at $z\simeq 0.5$ and their excess  brightness at $z > 0.5$, representing the transition
from an accelerated to a decelerated universe, are best appreciated by representing not the distance moduli 
but the distance moduli differences with respect to the empty Milne's universe in constant expansion. 

The grey dust models are represented as the distance moduli of a flat universe with 
$\Omega_{\mathrm{m0}}= 1$, $\Omega_{\mathrm{\Lambda 0}}=0$, with an atenuation obtained by

\be
\Delta m_{\rm dust} = -2.5 \log \Big \{ exp \big( C \int_{0}^{z} \rho_{\rm dust}(z)r(z)dz \big ) \Big \}
\ee
where $r(z)$ is the comoving distance travelled by the \s light, $\rho_{dust}(z)$ is given by equations
(1), (2) or (3), and $C$ is a constant used for dimensional purposes.

In the case of the SFR-based dust model proposed in the present work (3), it will be necessary to 
include a delay between dust formation and its incorporation into the IGM to contribute to the observed
extinction. To estimate the time interval in which the dust ejected could reach a homogeneous 
distribution in the intracluster space, we assume an expulsion velocity for the dust grains of 
around 1000 km s$^{-1}$, following Aguirre et al.\ (2001) and Shustov \& Vibe (1995). A delay of 
$\Delta t \approx 3\times 10^{9}$ years is obtained, implying that the homogeneity of the dust 
produced at $z \simeq 1$ is reached around a redshift of $0.3<z<0.4$. 

Figure 3 shows the different universe models, together with the weighted average data. From this figure
it can be concluded that the high-$z$ dust model can be discarded to a confidence level of $4 \sigma$ 
 for a homogeneous dust distribution. Also, it can be observed that the differences between the 
$\Omega_{\mathrm{m0}}= 0.31$, $\Omega_{\mathrm{\Lambda 0}}= 0.69$ universe and the replenishing dust 
model are almost negligible. Finally, the differences between our model with respect to the \s data are 
smaller than $1\sigma$. As discussed in \S 2, the advantage of the new model is that the dust 
distribution derives from current observational evidence. 

\begin{figure}[]
\centering
\includegraphics[width=9cm,height=7.5cm]{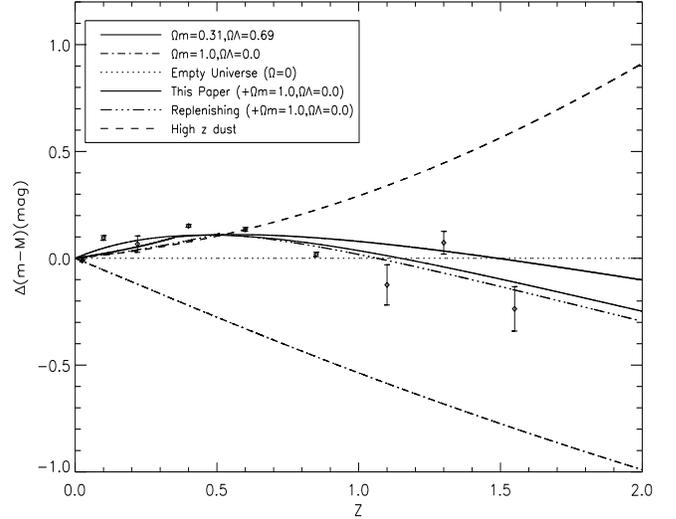}

\caption{\footnotesize{Averaged \s data and theoretical models showing the modulus difference  of
distances with respect to a universe in constant expansion ($\Omega = 0$). The high-$z$ dust 
model is rejected at $4 \sigma$ (dashed line). The best fit is given by an $\Omega_{mo}=0.31$, $\Omega_{\Lambda o}=0.69$  universe (solid thin line) 
with no  grey dust contribution. The SFR-based dust model represented takes into account a delay of 
2.7 Gy between the dust formation in the disk and its homogeneous distribution in the IGM.}}
\end{figure}

\begin{figure}[]
\centering

\includegraphics[width=9cm, height=7.5cm]{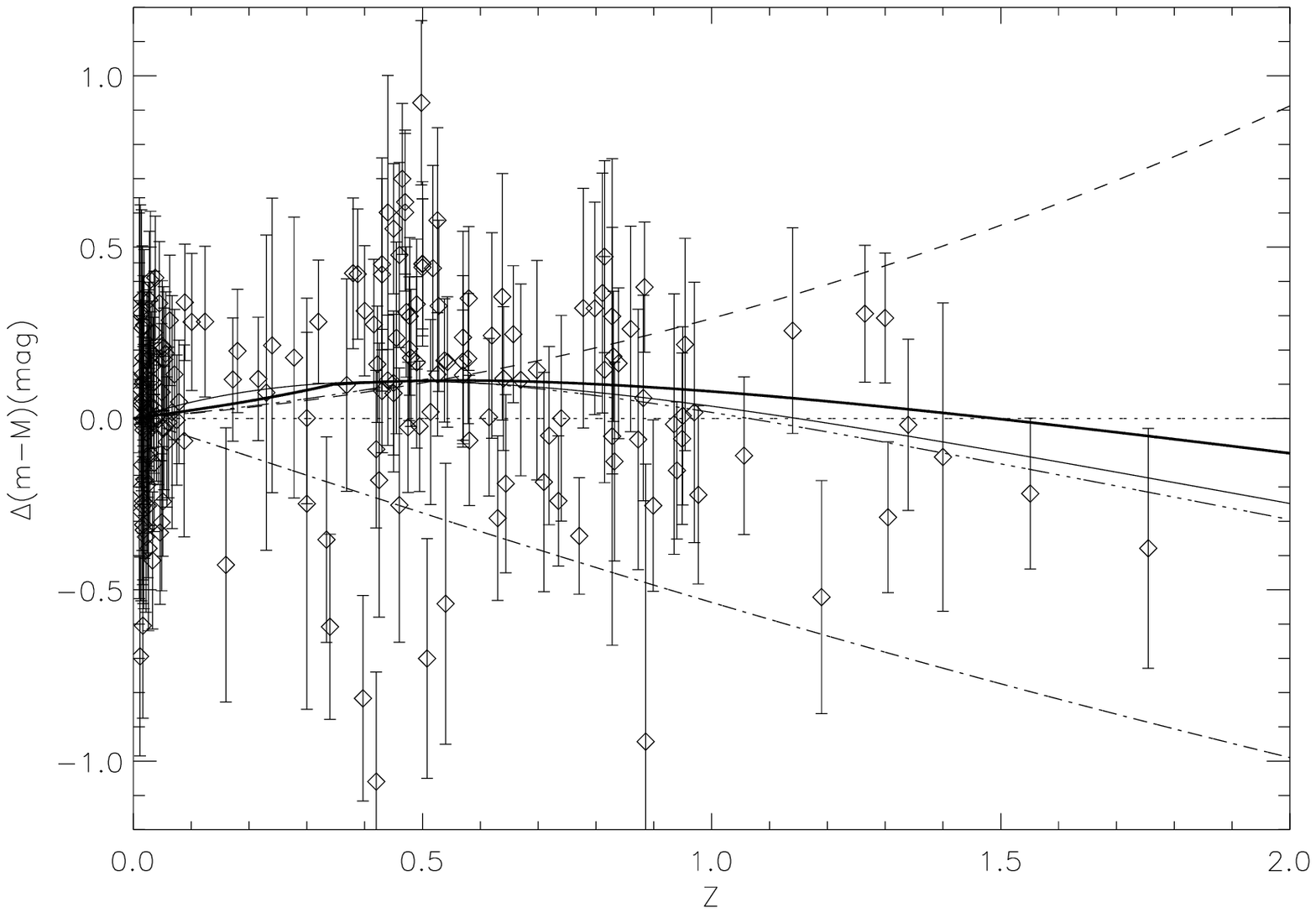}   
\includegraphics[width=9cm, height=7.5cm]{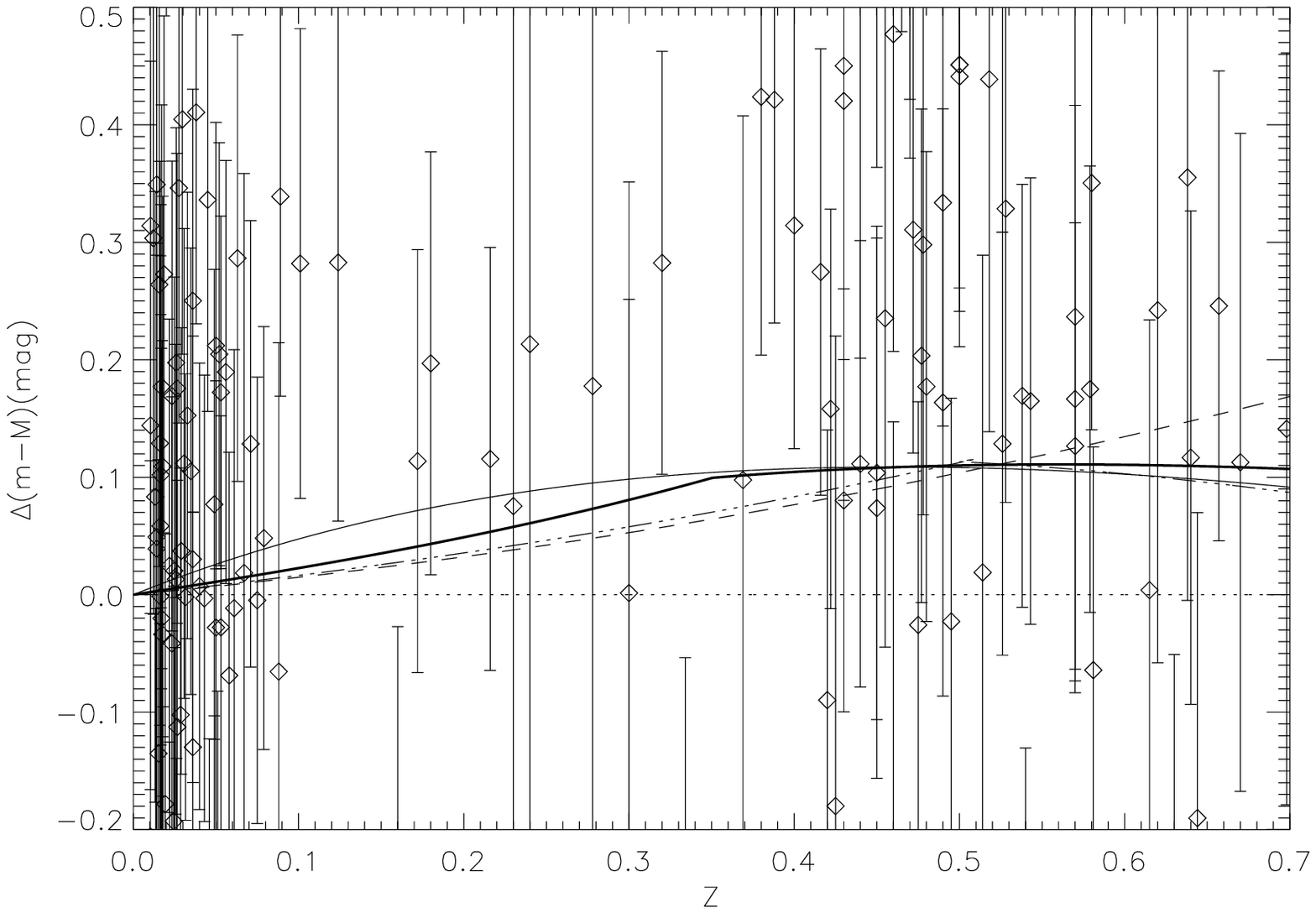}
\label{Figure 4}
\caption{\footnotesize{\emph{Top:} As in Figure 3, but representing the whole \s sample. 
The SFR-based distribution fits the data within the error bars of individual data.
\emph{Bottom:} Detail of the $0< z < 0.7$ region.}}
\end{figure}

In Figure 4 the entire sample of photometric data from \s is shown superposed on the theoretical 
models considered so far. It can be seen that the error bars of individual photometric data include 
both the SFR-based dust and the `replenishing' dust models. This model is quite flexible: the ambiguity in  current cosmic star formation history studies (expulsion velocities,
 delay times and SFR curve parameterization) does not have a big impact on our results, especially if the constants are empirically fitted as explained in Sect. 2.3.

In all these universe models (the grey dust ones), we have assumed a flat geometry with $\Omega_{m0}=1.0$, but this assumption 
contradicts the results of 2dF (Peacock et al.\ 2001). For this reason we have also developed an open universe 
($\Omega_{\mathrm{m0}}= 0.3$, $\Omega_{\mathrm{\Lambda 0}}=0.0$) with
SFR-based dust whose differences are less than $1\sigma$ compared
to the \s data (Figure 5 ). The implications of this open universe will be discussed in Sect. 4. 

\begin{figure}[]
\centering
\includegraphics[width=9cm,height=7.5cm]{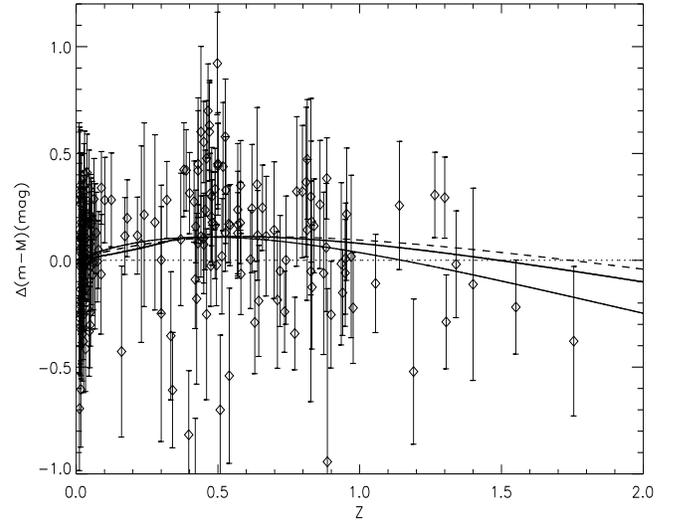} 
\label{Figure 5}
\caption{\footnotesize{Differences with respect to an empty universe. The flat (thick line) and open (dashed line)
SFR-based dust universe models are within the $1\sigma$ confidence level.}}
\end{figure}

\subsection{Further evidence against high-$z$ grey dust}

It might be argued that a non-homogeneous grey dust distribution could produce the deviation from 
the experimental data compared to Aguirre's model. According to this hypothesis, at high
redshifts only the \s whose line of sight to the observer passes through more voids (in other
words: whose ``beam'' is emptier) can be observed, since the others are too absorbed by the 
dust. In this way, distant \s are brighter than expected according to the high-$z$ dust model, lying somewhere 
between the high-$z$ grey dust and the $\Omega_{\mathrm{m0}}=1$ model, where the data points are 
located (Figure 4). 

To evaluate the feasibility of this hypothesis, the \s sample is searched for close pairs 
of \s with very different redshifts, in which the low $z$ ($z \simeq 0.5$) member of the pair 
is fainter than expected from the Milne model, while the high-$z$ one ($z \ge 0.9$) is not. Were 
such a pair to be found, it should be concluded that inhomogeneities in the high-$z$ grey dust model
cannot be deemed responsible for the observed \s distance modulus distribution.

Assuming a Robertson--Walker metric, the angle subtended by an object in a flat universe is

\be
\theta=\frac{D_{T}(z)}{D_{P}(z)},
\ee
where $D_{T}$ is the transversal distance and $D_{P}$ the proper distance. From this equation,
and assuming a cosmological model, the angle corresponding to a transversal distance can be
evaluated. 

To minimize the influence of photometric and model errors when applying this test, it is convenient 
to look for apparent \s pairs of very different redshifts, of which the \s at low $z$ show the 
highest brightness dimming. This maximum \s dimming is observed around $z=0.5$ (Riess et al.\ 1998, 2004; 
Perlmutter et al.\ 1999). Assuming a characteristic size of 30 Mpc for a void, the angle corresponding
to this distance at $z=0.5$ will represent the highest projected angular distance to search for a suitable 
pair to apply this test. This is a rather stringent requirement since a void size of 30 Mpc is smaller 
than those found by Einasto et al.\ (1994). From their data, an average diameter of around 
$\overline{\Phi_{\rm void}} \simeq 91 Mpc$ is deduced. In Table 1, the angles
corresponding to this distance at $z=0.5$ for different models of universe are shown. 

\begin{table}[h!]
\begin{center} 
\begin{tabular}{c c c}
\hline
\noalign{\smallskip}
$\Omega_{\mathrm{m0}}$ & $\Omega_{{\mathrm{\Lambda 0}}}$ & $\theta$ \\
\noalign{\smallskip}
\hline
\hline
\noalign{\smallskip}
0.3 & 0.0 & 0.9502\\
0.3 & 0.7 & 0.8451\\
1.0 & 0.0 & 1.0155\\
\noalign{\smallskip}
\hline
\end{tabular}
\label{Table 1}
\caption{\footnotesize{Values for the angle (in degrees) subtended by a $D_{T}$ of 30 Mpc at $z=0.5$ in 
different universes, with the assumed Hubble constant of 65 km s$^{-1}$ Mpc$^{-1}$}.}
\end{center}
\end{table}

With these hypotheses, the requirements to apply this test to supernovae pairs
are the following:

\begin{figure*}[]
\centering
\includegraphics[width=14cm,height=6cm]{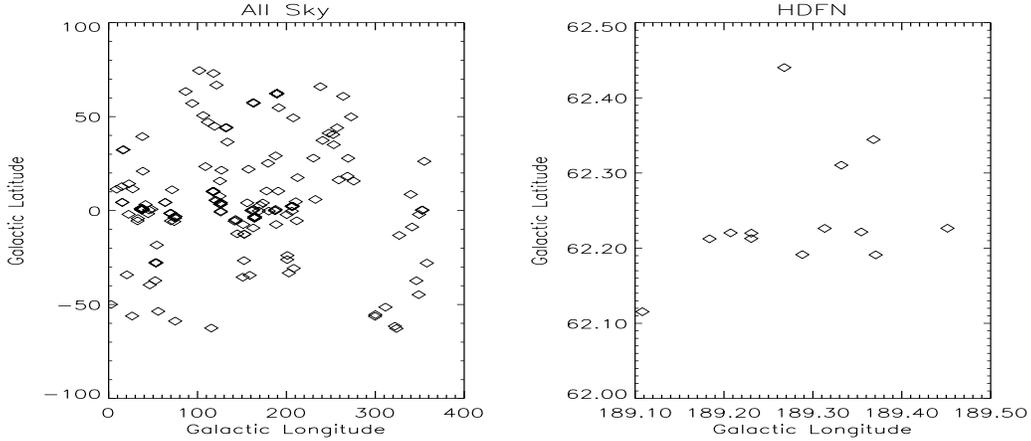}
\label{Figure 6} 
\caption{\footnotesize{Location on the sky of the \s sample. \emph{Left:} Position of all the supernovae 
in the sample. \emph{Right:} Detail of the {Hubble Deep Field North} region. The units on the axis 
are degrees.}}
\end{figure*}

\begin{enumerate}
\item The two members of the candidate pair  have to be in the ranges $0.35 < z < 0.65$ for the 
low $z$ and $z \ge 0.9$ for the high $z$, respectively. This requirement maximizes the difference in brightness of 
the pair members to reduce the contribution from photometric errors as much as possible.
\item For the same reason, the difference in redshift between the members of the pair must be $\Delta z \ge 0.4$. 
\item The angular distance on the sky between the two members of the pair must be lower than the value of $\simeq$1 degree assumed in Table 1  .
\item Both supernovae must be classified with the maximum confidence as type Ia. This requirement restricts the search to the \emph{Gold} sample.
\end{enumerate}

For such a small \s sample (157 \s) it might seem difficult to fulfil these requirements. 
Luckily, the systematic search for high-$z$ supernovae has centred on specific regions of the sky, 
mainly on the GOODS project HDFN and CDFS fields. The sky distribution of the \s sample is represented in 
Figure 6. The whole sample is shown in the left panel, while on the right the \s found in the HDFN are
shown, demonstrating the small distances that separate the \s in these deeply studied regions.

Table 2 summarises the results of our study for different cosmologies. Here, all the \s at $0.35\le z \le 0.65$ that are members of a pair are 
weaker than expected in an empty universe. A relatively large number
of pairs is found, considering that the whole sample is composed by only 186
SnIa.

\begin{table}[h!]
\begin{center} 
\begin{tabular}{ccccccc}
\hline
\noalign{\smallskip}
$\Omega_{\mathrm{m0}}$ & $\Omega_{\mathrm{\Lambda 0}}$ & $z_{1}$ & $z_{2}$ & Angle & 
Gold(A) & Silver(B)\\
\noalign{\smallskip}
\hline
\hline
\noalign{\smallskip}
0.3  & 0.0 & 0.35-0.65 & $\ge0.9$ & $\theta$ & 15 & 5 \\
0.3  & 0.0 & 0.35-0.65 & $\ge0.9$ & $\theta /2$ & 15 & 3 \\
0.3  & 0.7 & 0.35-0.65 & $\ge0.9$ & $\theta$ & 15 & 5 \\
0.3  & 0.7 & 0.35-0.65 & $\ge0.9$ & $\theta /2$ & 15 & 2 \\
1.0  & 0.0 & 0.35-0.65 & $\ge0.9$ & $\theta$ & 15 & 5 \\
1.0  & 0.0 & 0.35-0.65 & $\ge0.9$ & $\theta /2$ & 15 & 3 \\
\noalign{\smallskip}
\hline
\end{tabular}
\label{Table 2}
\caption{\footnotesize{Number of \s pairs found with an angular distance between members lower than 
the angle $\theta$ given by Table 1 for the corresponding cosmology. A: Pairs in which both
members belong to the \emph{Gold} sample. B: At least one of the members
belong to the \emph{Silver} sample.}}
\end{center}
\end{table}

\section{Conclusions}

With the available \s data we have calculated the best-fit dust-free universe model, obtaining 
$\Omega_{\mathrm{m0}}=0.31$, $\Omega_{\mathrm{\Lambda 0}}=0.69$. These results are very similar to those 
found in previous studies of luminosity distance to type Ia supernovae, leading to a reconciliation of the age of 
globular clusters with the age of the universe ($t_{0} \simeq 14.6 \times 10^{9}$ years, as derived from the
model, when assuming a cosmological constant-like dark energy, $\omega = -1$). Phantom energies can thus be 
avoided, as long as the mean values do not increase. However, the uncertainties in the age of globular
clusters are still too great to constrain the equation of state of dark energy.

A new grey dust model based on replenishing dust but including the observed comoving SFR 
density distribution has been derived, thereby providing an alternative model based on observational evidence. 
The extinction produced by this new SFR-based grey dust with $\Omega_{\mathrm{m0}}=1.0$ is very 
close to the $\Omega_{\mathrm{m0}}=0.31$, $\Omega_{\mathrm{\Lambda 0}}=0.69$ model with no extinction. 
Apart from the difficulties in providing the mass required for a flat universe and the age conflict
arising therefrom, the models would be distinguished only by future supernovae surveys at high $z$ with higher
photometric accuracy, such as those to be carried up with \emph{SNAP} (\emph{Supernova Acceleration Probe}) 
 (Aldering et al.\ 2002) and with the \emph{JWST} (\emph{James Webb Space Telescope}, Stockman et al.\ 1998),
 which could provide the evidence that will help to decide between a 
currently dark energy-dominated universe and a matter-dominated one. Moreover, the next generation \s surveys, 
with more accurate data, will be essential because the uncertainties in the current \s data, due to the 
contributions of approximations, photometric errors, luminosity maximum estimations at high $z$, etc., impose
 several restrictions on the validity of the cosmological results. Only when combined with other experiments such
  as CMB anisotropies or gravitational lenses is it possible to derive more stringent conclusions.

All the flat grey dust models presented here consider Euclidean geometry with an adimensional mass parameter, 
$\Omega_{\mathrm{m0}}=1$, a hypothesis against the data obtained so far (Peacock et al.\ 
2001). Even with the results of del Popolo (2003), who has suggested that current $\Omega_{\mathrm{m0}}$ 
values obtained from clusters of galaxies are underestimated by 20\%, a parameter $\Omega_{\mathrm{m0}}=1$ 
is excluded from the acceptable margins. For this reason we have included an open model $\Omega_{\mathrm{m0}}=0.3$,
 $\Omega_{\mathrm{\Lambda 0}}=0.0$ with our SFR-based grey dust attenuation. Although the confidence level of this
 model with respect to the \s data is below $1 \sigma$, it introduces a serious conflict with the interpretation of 
the WMAP results (Spergel et al.\ 2003). To reconcile this model with the CMB anisotropies experiment it is necessary to resort 
to exotic mechanisms, such as thermalization of the radiation due to the intracumular dust at high $z$.

The high-$z$ dust model is rejected out to $4 \sigma$ even if the grey dust presents a patchy 
distribution, as shown by the pair test devised here. However, most of the data at very high redshift 
corresponds to GOODS fields, which cover a small area on the sky (Figure 5). Pairs at other positions on
the sky combined with high-accuracy photometry would allow us to ascertain whether there is any significant
amount of grey dust in the universe. Nevertheless, the consistency of the
standard model and the weak basis for the alternative explanations makes the possibility of having an open model describing our universe difficult.

\begin{acknowledgements} 
We wish to thank G. A. Tammann, A. Tortosa Andreu, and A. M. Sánchez Quintana for helpful comments. 
This work was supported by the Spanish \emph{Plan Nacional de Astronom\'\i a y 
Astrof\'\i sica\/} under grant AYA2005-04149. This research made use of HZT \s photometric data, 
NASA's Astrophysics Data System (ADS), and the Nasa/IPAC Extragalactic Database (NED).
\end{acknowledgements}

\end{document}